\documentclass[epj]{svjour}

\usepackage{latexsym}
\usepackage{graphics}
\usepackage{graphicx} 
\usepackage{dcolumn}
\usepackage{epsfig}
\usepackage{amssymb} 
\usepackage{amsmath} 
\usepackage{rotating} 
\usepackage{color} 
\newcommand{\gammap}{\dot{\gamma}}
\newcommand{\re}{\mathcal{R}e}

\begin{document}

\title{An optical fiber based interferometer to measure velocity profiles in sheared complex fluids}

\author{Jean-Baptiste Salmon \and S\'ebastien Manneville \and Annie Colin \and Bernard Pouligny}
\institute{Centre de Recherche Paul Pascal, Avenue Schweitzer,
33600 Pessac, FRANCE}

\date{\today}

\abstract{We describe an optical fiber based interferometer
to measure velocity profiles in sheared complex fluids using Dynamic Light Scattering (DLS). 
After a review of the theoretical problem of DLS under shear, 
a detailed description of the setup is given. We outline the various experimental difficulties 
induced by refraction when using a Couette cell.
We also show that homodyne DLS is not well suited 
to measure quantitative velocity profiles in narrow-gap Couette geometries.
On the other hand, the heterodyne technique allows us to determine the velocity field 
inside the gap of a Couette cell. All the technical features of the setup, 
namely its spatial resolution ($\approx 50$--$100~\mu$m) and its temporal resolution ($\approx 1$~s per point,
$\approx 1$~min per profile) 
are discussed, as well as the calibration procedure with a Newtonian fluid.
As briefly shown on oil-in-water emulsions, such a setup permits one to record both velocity 
profiles and rheological data simultaneously.}


\authorrunning{J.-B. Salmon \textit{et al.}}
\titlerunning{A fiber based interferometer to measure velocity profiles in sheared fluids}
\maketitle

\section{Introduction}

The behaviour of complex fluids under flow raises important fundamental
problems and involves a very wide spectrum of possible applications.
A complex fluid is characterized by a {\it mesoscopic} scale,
located somewhere between the microscopic size of the molecules and
the macroscopic size of the sample \cite{Edimbourg:00,Larson:99}.
For instance, for oil-in-water emulsions,
this mesoscopic scale would be the size of an oil droplet.
The existence of this mesoscopic scale can make the flow of a complex fluid
considerably difficult to understand as compared to that of a
simple, viscous fluid. Indeed, most of complex fluids show
strong viscoelasticity and/or important nonlinear rheological
behaviours.

These behaviours arise from {\it coupling effects}
between the fluid microstructure and the flow. Usually, the presence
of a shear flow tends to order the fluid, for instance, by aligning
chains in a polymer solution and thus decreasing the viscosity 
(``shear thinning''). 
On the other hand, in a class of
concentrated systems recently called ``soft glassy materials'', such as concentrated
emulsions or colloidal glasses, yield stress and jamming phenomena are frequently encountered
\cite{Sollich:97,Viasnoff:02}.

Although the characterization of the {\it global} flow properties
remains a first essential step, much interest has grown in
performing {\it local} measurements on complex fluids under
shear. Indeed, in many cases, inhomogeneous flows are observed for
which the usual analysis in terms of pure shear flow is
not applicable and may be misleading. 

This paper is devoted to
local velocimetry in sheared complex fluids using Dynamic Light Scattering (DLS).
The need for local measurements is highlighted in the next Section through three
examples taken from recent works in complex fluids. 
We then review a few existing techniques that provide local measurements and explain
how they may or may not be suited to experiments in the concentric
cylinder geometry (referred to as ``Couette geometry'' in the following).
In Sec.~\ref{s.theory}, we give the basic theoretical principles of DLS under shear. 
In Sec.~\ref{s.explatex}, we present an experimental setup 
based on the use of single mode optical fibers.
Finally we show that the heterodyne geometry is better suited than the homodyne one
when the fluid under study is sheared in a Couette cell.

\section{The need for local measurements in the rheology of complex fluids}

\subsection{Examples of inhomogeneous flows in complex fluids}

Complicated rheological behaviours are observed in complex fluids
whose flow curves $\sigma(\dot{\gamma})$ are non-monotonic
($\sigma$ stands for the shear stress and $\gammap$ for the shear rate).
In such cases, above a critical shear rate, the stress may
decrease resulting in a mechanical instability of the flow
\cite{Spenley:93}. For controlled-shear-rate experiments, this type of
instability is expected to show up on the flow curve as a
stress plateau for $\gammap_1 < \gammap\ < \gammap_2$
separating two regions where the stress is an increasing
function of the shear rate. Such a signature is indeed observed
for various complex fluids such as wormlike micelle solutions \cite{Berret:97}
or some lyotropic lamellar phases \cite{Roux:93}. On the stress
plateau, the fluid is believed to phase-separate into two regions:
one with high viscosity and low shear rate flowing at $\gammap_1$ and
one with low viscosity and high shear rate $\gammap_2$. This
phenomenon is known as {\it shear banding}. In micellar
solutions, each ``band'' was shown to correspond to different fluid
structures, isotropic or nematic, using flow
birefringence \cite{Berret:97} or NMR spectroscopy \cite{Britton:97}.
The correspondence between the  branch flowing at $\gammap_2$ and
the nematic state is still under debate \cite{Fisher:01}.   

Thus, even though the geometry and the nature of the deformation
imposed on a complex fluid remain fairly simple (constant,
homogeneous force or velocity at the wall of a Couette cell for instance),
the response of the fluid may be highly nonlinear and
sometimes spatially inhomogeneous. In some extreme cases,
one not only observes bands of different structures and 
flow properties, but also large temporal fluctuations of the
global sample viscosity \cite{Wunenburger:01,Bandyopadhyay:01,Salmon:02}.
This last behaviour, recently coined ``{\it rheochaos}''
\cite{Cates:02}, is observed in the vicinity of the shear-induced
transitions described above. 
The occurence of such dynamical behaviours shows
that inhomogeneous flows of complex fluids
can exist both {\it spatially} and {\it temporally}
under simple shear, even in the absence of strong inertial
or elastic effects \cite{Muller:89,Groisman:00}.

A simpler but recurrent problem in rheology
is caused by {\it wall slip}.
Unlike simple fluids, mixtures under shear may
involve thin lubricating layers of low-viscosity fluid in
the vicinity of the cell walls. These lubricating
films absorb most of the viscous dissipation so that
the fluid in the bulk is submitted to a shear rate
smaller than that imposed globally. In emulsions, 
it is generally believed that apparent slip is due to
``wall depletion'', the fluid close to the walls containing
less oil droplets than in the bulk \cite{Barnes:95}. In classical
rheology, wall slip is
usually investigated by varying the cell characteristics
and/or the wall rugosity \cite{Larson:99}. In any case, it remains a
problem of considerable importance for industrial applications
and shows yet another occurence of inhomogeneous flows in
simple shear geometries.

\subsection{Experimental constraints in the Couette geometry}

Through the three examples of shear-banding, ``rheochaos,''
and wall slip, it clearly appears that {\it local} measurements are
crucial to understand the physics underlying the response of 
many complex fluids to deformation and flow. However, most
of the studies in this area still heavily rely on {\it global} data
recorded by standard rheometers imposing a torque (a
velocity resp.) on the axis of a moving cylindrical or conical part
and measuring its velocity (the torque resp.) once the fluid is
set into motion.

For more than two decades, great experimental effort has
been devoted to developping new methods for measuring
local quantities of rheological interest such as the local
stress, the velocity field, or the local velocity gradient \cite{Britton:97,Fisher:01,Sanyal:00_1,Welch:02}.
Ideally, such a technique should be implemented on standard rheometers.
On these apparatus, various geometries are used to induce a pure shear flow: 
(i) Couette cells in which the fluid is confined in the gap $e$ between two coaxial cylinders and where the
shear rate can be considered as homogeneous, provided
the ratio of the gap to the cylinder radii $e/R$ is not too
large,
(ii) cone-and-plate cells which provide
the best approximation of pure homogeneous shear flows,
and (iii) plate-and-plate cells where the shear rate increases
linearly with the radial position.

In the present paper, we will restrict ourselves to
shear flows generated in a Couette cell. 
This particular geometry imposes a few important constraints:

\begin{enumerate}
\item
The measurement technique has to be {\it non intrusive}.
Indeed, the very high sensitivity of the material
to local deformations precludes the use of a probe,
such as the hot wires or the hot films classically used in
hydrodynamics to record fluid velocities.
\item
The {\it thickness} $e=1$--3~mm and the {\it optical properties}
of the samples under investigation 
usually hinders direct optical access needed for 
techniques like Laser Doppler Velocimetry (LDV) or Fluorescence Correlation Spectroscopy (FCS):
most complex fluids are not transparent and
strongly scatter light. 
Also, in general, we are not able
to directly follow individual droplets in an emulsion
or fluorescent seeding particles in a lamellar phase.
Let us point out, however, that direct tracking of the system
is possible in the particular case of
two-dimensional foams \cite{Debregeas:01}
or when the fluid is transparent enough to allow Particle Imaging Velocimetry
(PIV) measurements such as those by Pine {\it et al.} in Ref.~\cite{Edimbourg:00}.
\item
Finally, the curved geometry of a Couette cell induces {\it refraction
effects} that can be relatively strong and make
optical techniques relying on light scattering tricky
to implement. 
\end{enumerate}

Moreover, a reliable measurement method should yield accurate,
reproducible measurements with good temporal and spatial resolutions.
Here, we shall call ``good'' a spatial resolution which provides, say,
20 measurement points accross the gap of our Couette cell,
which is typically $e=1$--3~mm wide. Any technique that is
able to distinguish volumes of fluid separated by 100~$\mu$m
and measure their respective velocities (or any other
local quantity) will thus have a good resolution. 

\subsection{NMR and DLS, two local measurements techniques}

Among the few quantitative local techniques that have proved
successful for complex fluids under flow are
(i) Nuclear Magnetic Resonance (NMR) and
(ii) Dynamic Light Scattering (DLS).

\subsubsection{Nuclear Magnetic Resonance}

Callaghan {\it et al.} have used NMR to image the velocity field
of sheared wormlike micelles in both cone-and-plate and Couette
($e=1$~mm) geometries \cite{Britton:97,Fisher:01}. Results in
the isotropic-nematic coexistence regime show a
strong correlation between shear bands and bands of different
structural order. NMR can resolve details down to
20~$\mu$m \cite{Britton:97,Fisher:01} and appears
as a very powerful tool for local measurements in complex fluids. 
However, the main drawback of this technique is its poor
temporal resolution: in Ref.~\cite{Fisher:01} the authors report on
a broad velocity distribution due to fluctuations of the
flow field but are not able to resolve these fluctuations in time.
Indeed, the accumulation time needed for one
profile is about an hour.

Moreover, this technique requires the costly production
and use of strong magnetic fields. Also, the flow cells
have to be made nonmagnetic. This leads to even more important
constraints on the design of the experiment.

\subsubsection{Dynamic Light Scattering}

Although subject to restrictions (2) and (3) listed above,
DLS is a low-cost technique easier to set up,
at least in the homodyne geometry. 
In the present work, these two restrictions can be partially
avoided by controlling the optical index of the complex
fluid under study. For instance by matching the indices of the 
aqueous phase and of the oil phase in an emulsion, we
are able to control the amount of scattered light and
avoid multiple scattering \cite{Salmon:02_3}

DLS is a common tool for probing spatio-temporal correlations
of the fluctuations of the dielectric permittivity in a material \cite{Berne:95}. 
With a realistic model of these fluctuations, DLS allows one to obtain
information on the dynamics of the fluctuations at length scale $\lambda$, 
the wavelength of incident light.
Such fluctuations may arise from different microscopic origins: if scattering particles 
are subject to Brownian motion then DLS yields a measurement of
the diffusion coefficient of these particles
\cite{Berne:95};
in lyotropic lamellar systems, fluctuations arise from undulating membranes and DLS may be a 
useful tool to characterize the elasticity of the surfactant bilayers \cite{Nallet:88}.
  
In the simple case of a shear flow, it has been shown that {\it homodyne} DLS may give information on
the local shear rate.
Such a technique has been used to measure shear rate
profiles of polymeric fluids in four-roll mill flows \cite{Sanyal:00_1,Wang:94}. 
However, this homodyne technique is not very well suited to the
Couette geometry as we will see in the following.
Ackerson and Clark \cite{Ackerson:81} have shown that {\it heterodyne} 
DLS was a powerful technique to measure local velocity and applied it
to colloidal crystals under Couette flow. With the same kind of heterodyne setup,
Gollub \textit{et al.} studied velocity profiles in Newtonian fluids
at the onset of the Taylor-Couette instability \cite{Gollub:74,Gollub:75}.
In spite of the above results, very few local measurements using heterodyne DLS have been
reported in the rheology literature. This is certainly due to the experimental difficulties
to set up an interferometer around a Couette cell.

\section{Theory of Dynamic Light Scattering in a shear flow}
\label{s.theory}

\subsection{Homodyne and heterodyne DLS}

\begin{figure}[htbp]
\begin{center}
\scalebox{1}{\includegraphics{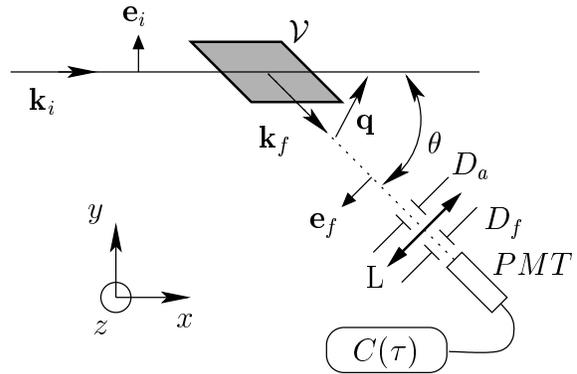}}
\caption{Geometry of homodyne DLS experiments. $PMT$ stands for photomultiplier tube.
Other symbols are defined in the text.} 
\label{montage_classique}
\end{center}
\end{figure}

The aim of this section is to give the basic theoretical principles of DLS under shear
which are necessary to reduce the optical information to local shear rate and local velocity measurements. 
In a typical DLS experiment as sketched in Fig.~\ref{montage_classique}, the sample is illuminated with
coherent polarized laser light. Scattered light is collected at an angle $\theta$
by a photomultiplier tube $PMT$.

The optical device which collects light is composed of 
a field diaphragm $D_f$ and a lens $L$. The scattering volume $\mathcal{V}$ is the intersection of the 
incident beam with the image of $D_f$ through $L$. The aperture diaphragm $D_a$ allows one to reduce 
the range of collected angles $\theta$ (it controls the size of the photodetection area). 
The auto-correlation function
$C(\tau) = <\!i(t)i(t+\tau)\!>$ of the scattered intensity $i(t)$ is measured by an electronic correlator
as a function of the time lag $\tau$.
$\mathbf{e_i}$ is the direction of the polarization of the incident beam and $\mathbf{e_f}$ is
the direction of the analyzer between the scattering volume and the $PMT$.
  
\subsubsection{Homodyne DLS}

When the dielectric permittivity tensor $\overline{\overline{\epsilon}}(\mathbf{r},t)$ of the material is 
slightly perturbated, i.e. when $\overline{\overline{\epsilon}}(\mathbf{r},t) = \epsilon \mathcal{I} + 
\overline{\overline{\delta\epsilon}}(\mathbf{r},t)$, and
$ \delta\epsilon \ll \epsilon $ 
($\mathcal{I}$ is the unity tensor),
the magnitude of the electric field scattered at angle $\theta$ 
is \cite{Berne:95}: 
$$E_s(t) \propto \widehat{\delta \epsilon}_{if}(\mathbf{q},t)\, , $$
where $ \mathbf{q} = \mathbf{k}_i - \mathbf{k}_f $ is the scattering vector, whose magnitude is given by
$q = (4 \pi n /\lambda) \sin(\theta/2) $, 
and $\widehat{\delta \epsilon}_{if}(\mathbf{q},t)$ is the 
$i f$-component of the Fourier transform of the dielectric permittivity tensor of the sample restricted to
the scattering volume $\mathcal{V}$:
$$ \widehat{\delta \epsilon}_{if}(\mathbf{q},t) = 
\int_{\mathcal{V}} d^3\mathbf{r} e^{i\mathbf{q}\cdot\mathbf{r}} 
\mathbf{e_i} \overline{\overline{\delta\epsilon}}(\mathbf{r},t) \mathbf{e_f}\, . $$
In the following, we will focus on the $z z$-component of the tensor 
$\overline{\overline{\delta\epsilon}}(\mathbf{r},t)$, and we write for simplicity:
$\delta\epsilon(\mathbf{r},t) = \delta\epsilon_{zz}(\mathbf{r},t)$.    

Measuring the intensity scattered at angle $\theta$ allows one to 
determine the spatial fluctuations of the dielectric
permittivity at length scale $q^{-1}$. 
In DLS, temporal fluctuations are studied through
the correlation function $C(\tau) = <\!i(t)i(t+\tau)\!>$. If $\delta\epsilon(\mathbf{r},t)$ is a Gaussian 
random variable, one can show that \cite{Berne:95}: 
\begin{equation}
C(\tau) =  A + B|g^{(1)}(\tau)|^2\, ,
\label{chomo}
\end{equation}
where
$$  g^{(1)}(\tau) = 
<\!\widehat{\delta\epsilon}(\mathbf{q},t) \, \widehat{\delta\epsilon}^*\!(\mathbf{q},t+\tau)\!>\, .$$
$A$ and $B$ are two constants that depend on the signal-to-noise ratio.
$g^{(1)}(\tau)$ is the auto-correlation of the Fourier transform of the 
fluctuations of $\delta\epsilon(\mathbf{r},t)$ at length scale $q^{-1}$ and time $\tau$.
A device measuring $|g^{(1)}(\tau)|^2$ is called a \textit{homodyne} DLS setup.

\subsubsection{Heterodyne DLS}
In the case where the temporal fluctuations of $\delta\epsilon(\mathbf{r},t)$ 
are non-Gaussian or when $g^{(1)}(\tau)$ 
contains phase terms, homodyne DLS is not sufficient. To access the real part of $g^{(1)}(\tau)$,
one has to use DLS in \textit{heterodyne} mode.
This method consists in measuring the signal resulting from the interference
between the scattered electric field $\mathbf{E}_s$ and a \textit{local oscillator} $\mathbf{E}_{\scriptscriptstyle LO}$ 
taken from the incident beam, whose magnitude
is much greater than the scattered electric field.  
In this case, one can show that the autocorrelation function of $\mathbf{E}_s+\mathbf{E}_{\scriptscriptstyle LO}$ 
satisfies  \cite{Berne:95}:
\begin{equation}
C(\tau) =  A' + B'\re\left(g^{(1)}(\tau)\right)\, ,
\label{chete}
\end{equation}
where $A'$ and $B'$ are two constants that depend on the signal-to-noise ratio.
\subsection{A simple model for $g^{(1)}(\tau)$  in a shear flow}

\subsubsection{Physical considerations}

\label{s.simplemodel}

\begin{figure}[htbp]
\begin{center}
\scalebox{1}{\includegraphics{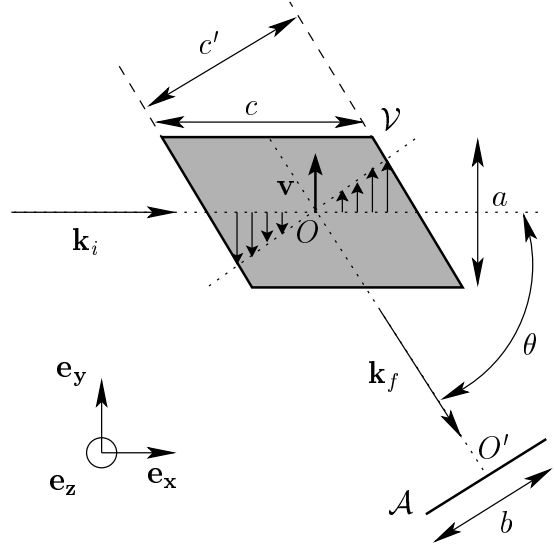}}
\caption{Geometry of the scattering volume in a shear flow. $\mathcal{V}$ stands for the scattering volume and
 $\mathcal{A}$ for the photocathode area.}
\label{geometryflow}
\end{center}
\end{figure}

Let us now turn to the case when the scattering volume is submitted to a shear flow, as shown in Fig.~\ref{geometryflow}.
The theoretical problem of DLS in a shear flow has been extensively studied 
\cite{Wang:94,Ackerson:81,Fuller:80,Maloy:92}.
The main result of these works is that \textit{homodyne} DLS allows one to measure the local shear rate 
through decorrelation terms of geometrical nature.

On the other hand, \textit{heterodyne} DLS yields an estimate of the mean
velocity in the scattering volume using the
well-known Doppler effect. Indeed, since scattering particles are moving at 
velocity $\mathbf{v}$, the frequency of the scattered electric field is 
Doppler shifted by the quantity $\mathbf{q}\!\cdot\!\mathbf{v}$. 
It is straightforward to show that $g^{(1)}(\tau) = F(\tau)\,\exp(i\mathbf{q}\!\cdot\!\mathbf{v} \tau)$, where
$F(\tau)$ is some decreasing geometrical factor. 
Since homodyne DLS measures the modulus of $g^{(1)}(\tau)$, the local velocity can only be accessed by using the
heterodyne configuration.  

The aim of this section is to give some simple physical ideas that help to 
understand the correlation functions measured with heterodyne DLS. 
We then present more detailed equations which will be solved numerically in Sec.~\ref{s.explatex}.

The geometry of DLS under shear is shown in Fig.~\ref{geometryflow}. The incident beam is normal to the velocity ($y$)
and to the vorticity ($z$) directions. 
In a shear flow, one expects the following invariance rule for the correlation of the fluctuations 
\cite{Ackerson:81}:
\begin{align}
\Gamma(\mathbf{r},t,\mathbf{r'},t')  \widehat{=} &  
<\!\delta\epsilon(\mathbf{r},t) \delta\epsilon(\mathbf{r'},t')\!> 
\notag 
\\  = &
\, \Gamma(\mathbf{r'}-\mathbf{r} - (\mathbf{v} + \dot{\gamma}(\mathbf{r}.\mathbf{e}_x)\mathbf{e}_y)(t'-t), t'-t)\, . 
\label{invariance}
\end{align}
Thus, in the presence of a shear flow,
the physical properties
in the scattering volume are no longer invariant under spatial translations.  
Hence the spatial coherence of the scattered electric field is modified \cite{Maloy:92}.
From Eq.~(\ref{invariance}), one
can show that: 
\begin{equation}
 <\!\widehat{\delta\epsilon}(\mathbf{q},t) \widehat{\delta\epsilon}(\mathbf{-q'},t+\tau)\!> \neq 0\, ,
\end{equation}
only if the following condition is satisfied:
\begin{equation}
\mathbf{q'} = \mathbf{q} +  \dot{\gamma}\tau(\mathbf{q}.\mathbf{e}_y)\mathbf{e}_x\, .
\label{condition}
\end{equation}

In fact, this relation is strictly true only for an infinite scattering volume. 
In practice, $\mathcal{V}$ corresponds to the intersection of two infinite cylinders.
Therefore, the scattering volume has dimensions of the order of $a\times a \times c$ (cf. Fig.~\ref{geometryflow}), 
and the scattering vectors $\mathbf{q}$ are only defined up to the precision $\delta q_x = \pm 2\pi/c$ and
$\delta q_y = \delta q_z = \pm 2\pi/a$. 
Due to the finiteness of the scattering volume, light scattered at a distance 
$R$ corresponds to \textit{speckle grains} 
and is spatially coherent in small \textit{coherence areas} of typical size $\lambda R/c$.
In the experiments, the aperture diaphragm $D_a$ collects a small range of angles $\theta$ and therefore 
a small range of $\mathbf{q}$'s. In this more realistic case, the correlation function measured with heterodyne DLS 
reads:
\begin{equation}
g^{(1)}(\tau) = \sum_{\mathbf{q},\mathbf{q'}\in \mathcal{A}} <\!\widehat{\delta\epsilon}(\mathbf{q},t) \widehat{\delta\epsilon}(\mathbf{-q'},t+\tau)\!>\, ,
\label{C_tau_photocathode_etendue}
\end{equation}
where the sum is calculated over the coherence areas which are collected on the $PMT$.
In the presence of shear, due to the condition of Eq.~(\ref{condition}), one can see that 
coherence areas are moving in the $x$-direction. This leads to decorrelation
when $\|\mathbf{q'}-\mathbf{q}\| > \delta q_x $ i.e. when $\tau > \tau_{\dot{\gamma}} = 2 \pi/(c \dot{\gamma} q_y)$.

Another decorrelation time arises from the \textit{finiteness} of the photocathode area. 
Indeed, speckle grains move out of the 
photodetection area due to the spatial coherence condition of Eq.~(\ref{condition}). The separation $\delta R$ 
at a time $\tau$ between two coherence areas which contribute to $C(\tau)$ is of the order of
$\delta R \approx \|\mathbf{q'}-\mathbf{q}\| R/k_f \approx \dot{\gamma} R \tau$. Thus, for times
$\tau > \tau_{\mathcal{A}} = b/(\dot{\gamma} R)$, this term leads to the decorrelation of $C(\tau)$
because the two coherence areas are separated by more than the characteristic length $b$ of the photocathode area.     

Finally there is a third decorrelation time due to the \textit{transit time} $\tau_t$ 
of the particles through
the scattering volume $\mathcal{V}$. This transit time is of the order of $\tau_t \approx a/v$, 
where $v$ is the mean velocity in the scattering volume. In a lot of experimental situations however
one has: $\tau_t \gg \tau_{\dot{\gamma}}$ and $\tau_t \gg \tau_{\mathcal{A}}$
so that
this decorrelation cannot be observed. Physical systems may also display
intrinsic dynamics such as Brownian motion.
Such dynamics leads to decorrelation but in most systems the decorrelation
times associated to those dynamics are very 
large compared to the geometrical times discussed
above and cannot be observed \cite{Ackerson:81}.   

From these physical considerations, we can assess that $g^{(1)}(\tau)$ takes the following form:
\begin{equation}
g^{(1)}(\tau) = e^{i\mathbf{q}\cdot\mathbf{v}\tau} F(\tau_{\mathcal{A}},\tau_{\dot{\gamma}},\tau)
\label{g1tau_general}
\end{equation}
 
\subsubsection{Derivation of $F(\tau_{\mathcal{A}},\tau_{\dot{\gamma}},\tau)$}

\begin{figure}[htbp]
\begin{center}
\scalebox{1}{\includegraphics{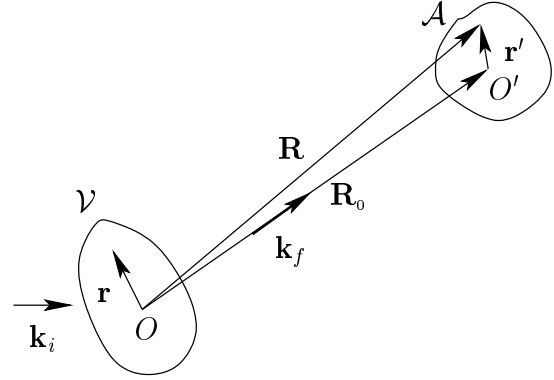}}
\caption{Typical configuration for light scattering in realistic conditions. 
$\mathcal{V}$ is the scattering volume, $\mathcal{A}$ the photocathode area}
\label{difftheorique}
\end{center}
\end{figure}

We now present a detailed derivation of the heterodyne correlation
function under shear. In order to
estimate $C(\tau)$  in realistic conditions, one must take into account
 the finiteness of the scattering volume $\mathcal{V}$ and of
the photocathode area $\mathcal{A}$ as sketched in Fig.~\ref{difftheorique}. 
One can show that the magnitude of the electric field scattered at point $\mathbf{R} = \mathbf{R_{\scriptscriptstyle 0}} + \mathbf{r'}$
from the scattering volume $\mathcal{V}$ is given by \cite{Berne:95}:
\begin{equation}
E_s(\mathbf{R},t) = \frac{k_f^2 E_{\scriptscriptstyle 0}e^{i(k_iR_{\scriptscriptstyle 0} -\omega t)}}{4 
\pi \epsilon_{\scriptscriptstyle 0} R_{\scriptscriptstyle 0} } 
e^{i\mathbf{k}_f.\mathbf{r'}}
\widehat{\delta\epsilon}\left(\mathbf{q} + \frac{k_f}{R_{\scriptscriptstyle 0}}\mathbf{r'},t\right) \, ,   
\label{E_s}
\end{equation}
with $\mathbf{q} = \mathbf{k}_f - \mathbf{k}_i$ and assuming that 
$R \gg r$ and $R \gg r'$. $R_{\scriptscriptstyle 0}$ is the distance between the center $O'$ of the photocathode area 
$\mathcal{A}$ and the center $O$ of the scattering volume $\mathcal{V}$.

In heterodyne DLS, the normalized correlation function takes the following form:
\begin{equation}
C(\tau) \propto \mathcal{R}e\left(\iint_{\mathcal{A}} <\!E_s(\mathbf{R'},t) 
E_s^*(\mathbf{R''},t+\tau)\!>\right) \, . 
\end{equation}
Using Eq.~(\ref{E_s}), one gets: 
\begin{align}
C(\tau) \propto  \mathcal{R}e & \iint_{\mathcal{A}} d\mathbf{r'}d\mathbf{r''} e^{i\mathbf{k}_f\cdot
(\mathbf{r'}-\mathbf{r''})} \notag \\
& \left<\!\widehat{\delta\epsilon}\left(\mathbf{q} + \frac{k_f}{R_{\scriptscriptstyle 0}}\mathbf{r'},t\right)
\widehat{\delta\epsilon}^*\left(\mathbf{q} + \frac{k_f}{R_{\scriptscriptstyle 0}}\mathbf{r''},t+\tau\right)\!\right> 
 \, .
\label{Ctau}
\end{align}

Including the effect of shear requires a model for the dielectric
permittivity of the material. A simple approach consists in considering isotropic, 
independent, point-like scatterers moving in a shear flow. 
In such a case, 
$\delta\epsilon(\mathbf{r},t) = \sum_{i=1}^N \delta(\mathbf{r}-\mathbf{r_i}(t))$ 
and
$\mathbf{r_i}(t+\tau) = \mathbf{r_i}(t) + \mathbf{v}\tau +\dot{\gamma}x \mathbf{e}_y\tau $ 
(see Fig.~\ref{geometryflow}). Eq.~(\ref{Ctau}) can then be rewritten as~:

\begin{align}
C(\tau) \propto \mathcal{R}e & \iint_{\mathcal{A}} d\mathbf{r'}d\mathbf{r''} e^{i\mathbf{k}_f\cdot(\mathbf{r'}-\mathbf{r''})}
\notag \\ & \int_{\mathcal{V}}d\mathbf{r}\, e^{i \mathbf{q}\cdot(\mathbf{v} +\dot{\gamma}x \mathbf{e}_y)\tau} 
e^{i (k_f/R_{\scriptscriptstyle 0})
[ \mathbf{r}\cdot(\mathbf{r'}-\mathbf{r''})+\mathbf{r'}\cdot(\mathbf{v} 
+ \dot{\gamma}x \mathbf{e}_y)\tau]} \, .
\label{C_tau_developpe}
\end{align}
To proceed further in the discussion, the shape of the scattering volume and of the photocathode area have to be specified, 
so that Eq.~(\ref{C_tau_developpe}) can be solved numerically. 

\section{DLS experiments in a Couette flow}
\label{s.explatex}

\subsection{Description of the experimental device}
\label{s.setup}

\begin{figure}[htbp]
\begin{center}
\scalebox{1}{\includegraphics{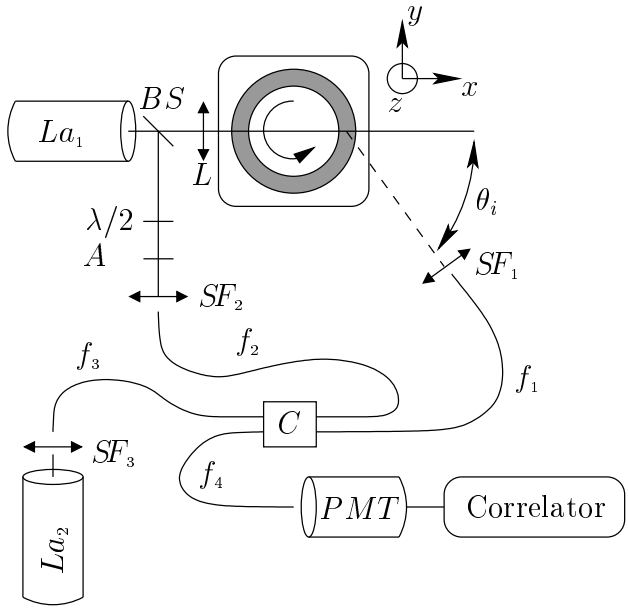}}
\caption{Heterodyne DLS setup. $La$ denotes lasers, $BS$ a beam splitter, $SF$ spatial filters, and $C$ the device
coupling optical fibers $f$}
\label{setup}
\end{center}
\end{figure}

As discussed in the Introduction, one of the major challenges in the study of complex fluid flows is 
to measure simultaneoulsly the global rheology of a material under shear 
and the velocity profile inside the fluid.
In order to perform such experiments we have developed the setup
sketched in Fig.~\ref{setup}, where heterodyne DLS has been installed
around a classical rheometer (TA Instruments AR 1000).

In order to probe the rheological properties of our complex fluids,
we use a transparent Mooney-Couette cell
whose geometrical characteristics are given in the next paragraph.
The temperature of the sample is controlled within $\pm\,0.1^{\circ}$C using a water circulation around the cell.
The rheometer sits on a mechanical table whose displacements are controlled by a computer.
Three mechanical actuators allow us to move the rheometer in the $x$, $y$, and
$z$ directions with a precision of $1~\mu$m. 

A $He$-$Ne$ polarized laser beam $La_{\scriptscriptstyle 1}$ ($40$~mW, $\lambda = 632.8$~nm) 
is directed through the cell and is focussed inside the gap
by a lens $L$ ($f=10$~cm). The incident beam is polarized along the $z$ direction. 
A spatial filter $SF_{\scriptscriptstyle 1}$, composed of 
a microscope lens and a single mode optical
fiber $f_{\scriptscriptstyle 1}$ (core diameter $ d= 4~\mu$m, IDIL) collects the  
electric field scattered at an imposed angle $\theta_i$
by the sample lying in the gap.

In order to perform heterodyne measurements, a local oscillator
is obtained from the incident beam by
a beam splitter $BS$ and a
second spatial filter $SF_{\scriptscriptstyle 2}$ 
which directs light into single mode fiber $f_{\scriptscriptstyle 2}$.
The interference between this local oscillator  and the 
scattered electric field is achieved by a single mode coupling device $C$ 
which connects fibers $f_{\scriptscriptstyle 1}$ and $f_{\scriptscriptstyle 2}$. 
This allows for great flexibility when choosing the scattering angle.
From the output of $C$, the electric field arising from the interference propagates through 
single mode fiber $f_{\scriptscriptstyle 4}$ to a photomultiplier tube $PMT$ (Thorn EMI Type 9868), 
which is connected
to an electronic correlator (Malvern Series 7032 with 64 channels). 

Between $SF_{\scriptscriptstyle 2}$ and beam splitter $BS$, an attenuator
$A$ avoids a too high illumination of the photomultiplier
tube. A retardation plate ($\lambda/2$ plate) 
permits us to change the polarization of the light entering
$f_{\scriptscriptstyle 2}$.
The axis of the $\lambda/2$ plate is tuned in such a way
that the two electric fields entering $C$ have the same polarization. Thus,
the interference between the scattered electric field and
the local oscillator measured with the $PMT$ is optimal.

To access the geometrical features of the scattering volume $\mathcal{V}$, we  
can turn on a second laser $La_{\scriptscriptstyle 2}$ ($He$-$Ne$, $10$~mW), 
which is coupled to single mode fiber $f_{\scriptscriptstyle 3}$ via  
spatial filter $SF_{\scriptscriptstyle 3}$. 
Light from $La_{\scriptscriptstyle 2}$ propagates through 
$f_{\scriptscriptstyle 3}$, 
then through $f_{\scriptscriptstyle 1}$ and
goes out via $SF_{\scriptscriptstyle 1}$. Therefore we get a direct 
visualization of the scattering volume $\mathcal{V}$  as the
intersection of the incident beam with 
the beam from $La_{\scriptscriptstyle 2}$.

\subsection{Design of the Couette cell and refraction effects}
\label{refraction}
\begin{figure}[htbp]
\begin{center}
\scalebox{1}{\includegraphics{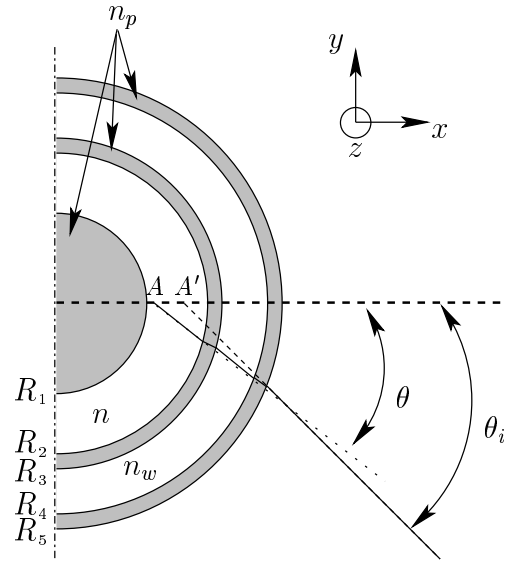}}
\caption{Experimental Couette cell. Plexiglas (optical index $n_p$) is indicated in gray. The sample (optical index $n$) lies between $R_{\scriptscriptstyle 1}$ and $R_{\scriptscriptstyle 2}$
and the thermostated water (optical index $n_w$) between $R_{\scriptscriptstyle 3}$ and $R_{\scriptscriptstyle 4}$.}
\label{couette}
\end{center}
\end{figure}
In the experimental setup described in Sec.~\ref{s.setup}, one of the major constraints is the use
of a thermostated Couette cell.
Indeed, our  home-made cell sketched in Fig.~\ref{couette} is composed
of four cylindrical interfaces. These interfaces separate the fluid under study from the water 
used for temperature control. 
This Couette cell thus acts as a cylindrical lens which leads to various optical
complications: 
(i) the angle $\theta_i$ imposed by the operator differs 
from the actual scattering angle $\theta$, and
(ii) the scattering volume lies at point $A$ instead of point $A'$ (see Fig.~\ref{couette}). This 
leads to intricacies in the experimental determination of the positions of the stator and 
the rotor. Moreover when the Couette cell is moved in the $x$ direction by a quantity $\delta x$, 
the real displacement  $\delta x'$ of the scattering volume slightly differs from $\delta x$.   

All these problems due to \textit{refraction} must be taken into account in order to get reliable velocity profiles.
To explicit the different relationships between $\theta$, $\theta_i$, $\delta x$, and  $\delta x'$,
we simply write the Snell-Descartes laws for the four interfaces. The geometrical parameters are as follows:
$R_{\scriptscriptstyle 1} = 22, R_{\scriptscriptstyle 2} = 25, R_{\scriptscriptstyle 3} = 27.6, 
R_{\scriptscriptstyle 4} = 30.5 $ and $R_{\scriptscriptstyle 5} =35.6$~mm.
Hence, the aspect ratio of our Couette cell is:
$(R_{\scriptscriptstyle 2}-R_{\scriptscriptstyle 1})/R_{\scriptscriptstyle 2} = 0.12$.
All the fluids studied in this paper have the same optical
index $n=1.40$. The optical indices of Plexiglas and  water are $n_p=1.47$ and $n_w = 1.33$ respectively.
The imposed angle  is $\theta_i = 35^{\circ}$.
With such parameters, Snell-Descartes laws lead to $\delta x' = f_x \delta x$
with $f_x \approx 1.13$ and $ \theta = f_{\theta} \theta_i$ with $f_\theta \approx 0.79$. 
We checked that $f_x$ does not depend on the position in the gap up to a very good approximation.
However $\theta$ depends slightly on $x$ via $f_\theta$: its relative variation is
$\delta \theta / \theta \approx 3~\%$ across the cell gap. 

This last point is very important because the experimentally measured quantity,
the Doppler shift $\mathbf{q}\!\cdot\!\mathbf{v}$,
depends on the scattering angle.
When scanning the width of the gap during an experiment, this small variation of $\theta$
leads to the following intrinsic incertitude on the velocity $\mathbf{v}$:
$\delta v / v \approx 3~\%$.   
Let us notice however that this uncertainty strongly depends on the gap width: 
in a narrower gap, for instance $e=1$~mm, $\delta v / v \approx 1~\%$. 

In the next paragraph we show that the use
of single mode fibers leads to 
a simplification of Eq.~(\ref{C_tau_developpe}) and we present measurements of
the shape of the scattering volume using
laser $La_{\scriptscriptstyle 2}$. Such measurements will allow us to compute the
theoretical correlation function.  

\subsection{Numerical derivation of the heterodyne correlation function under shear}
\label{s.numerical_integration}

\subsubsection{A simplification due to single mode fibers}

A single mode fiber intrinsically
collects coherent light only. Therefore the number of coherence areas detected by single mode fiber 
$f_{\scriptscriptstyle 1}$
is one: the size of the photocathode area does not play 
any role in our case. This feature is consistent with the 
experimental results presented in Sec.~\ref{s.expfeat} where we will stress the fact that the only relevant 
geometrical time is $\tau_{\gammap}$.
According to this simplification, Eq.~(\ref{C_tau_developpe}) can be rewritten as:
\begin{align}
C(\tau) & \propto  \cos(\mathbf{q}\!\cdot\!\mathbf{v})
\int_{\mathcal{V}} e^{i\dot{\gamma}x q_y\tau} \, . 
\label{tf_volume_diffusant}
\end{align}   
To proceed further in the integration of Eq.~(\ref{tf_volume_diffusant}), 
one has to know the exact shape of the scattering volume.

\subsubsection{Shape of the scattering volume}
\label{s.size}

The geometry of the scattering volume 
depends on the field diaphragm $D_f$, the focal length of the lens used
in spatial filter $SF_{\scriptscriptstyle 1}$ and the distance $R$ between $SF_{\scriptscriptstyle 1}$ and the scattering volume $\mathcal{V}$.  

\begin{figure}[htbp]
\begin{center}
\scalebox{0.4}{\includegraphics{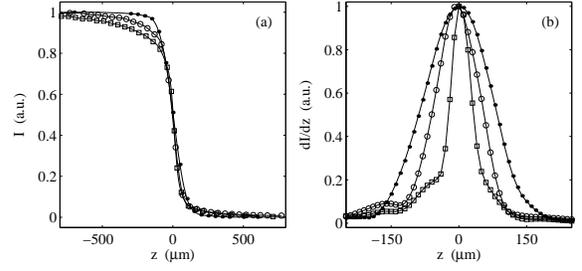}}
\end{center}
\caption{(a) Intensity collected on a photodiode by
moving a sharp blade across the beam collecting the scattered
light and visualized by turning on laser La$_2$ (see Fig.~\ref{setup}).
(b) Intensity profile across the beam collecting the scattered light.
The focal length of the microscope lens of spatial filter $SF_{\scriptscriptstyle 1}$
is 2.9~mm ($\bullet$), 5.5~mm
($\circ$), and 11~mm ($\Box$).}
\label{couteau}
\end{figure}

To determine the size of $\mathcal{V}$, we image $\mathcal{V}$
using laser $La_2$ as explained in Sec.~\ref{s.setup}.
With a sharp blade moving across the beam and perpendicular to the beam, 
we measured the transmitted intensity for different 
heights $z$ of the blade (see Fig.~\ref{couteau}(a)).

From this measurement, the intensity profiles of the beam 
are easily obtained by differentiating the previous signals (see Fig.~\ref{couteau}(b)). 
In Fig.~\ref{couteau}, the
focal length of spatial filter $SF_{\scriptscriptstyle 1}$ was varied and the intensity profiles were 
measured at the distance $R=13$~cm from $SF_{\scriptscriptstyle 1}$ after the image of single mode fiber 
$f_{\scriptscriptstyle 1}$ was focussed 
on the blade (see below). Such a method allows us to estimate the typical size $c'$ of the scattering volume.
We chose to define $c'$ as the half-width of the intensity profiles 
at the relative intensity $I/I_{\scriptscriptstyle 0} = e^{-2}$.
As shown in Fig.~\ref{couteau}(b), the minimum size 
is obtained for the focal length $f=11$~mm: $c' \approx 40~\mu$m.
This is consistent with the geometrical 
magnification factor $\gamma \approx R/f \approx 10$
which yields $c' \approx \gamma d \approx 40~\mu$m, where $d = 4~\mu$m is the core diameter of the fiber
and $R = 13$~cm.

\begin{figure}[htbp]
\begin{center}
\scalebox{0.7}{\includegraphics{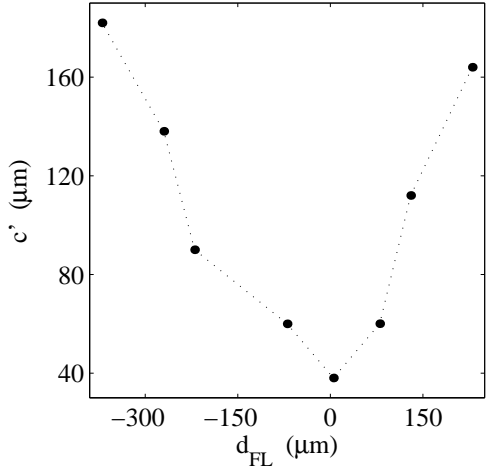}}
\end{center}
\caption{Width of the scattering volume $c'$
 measured from the intensity profiles for different ``defocusing'' distances $d_{\scriptscriptstyle FL}.$
The focal length of $SF_{\scriptscriptstyle 1}$
is 11~mm.}

\label{couteau2}
\end{figure}

To make sure that the image of single mode
fiber $f_{\scriptscriptstyle 1}$ is located exactly on the blade, one has to
measure the width of the beam at the position of the blade for different \textit{defocusing}  
distances $d_{\scriptscriptstyle FL}$ between the lens 
and single mode fiber $f_{\scriptscriptstyle 1}$.   
In Fig.~\ref{couteau2}, we plotted the characteristic length $c'$ measured
for different distances $d_{\scriptscriptstyle FL}$ 
with the lens of focal length $f=11$~mm.
This curve shows a minimum at the distance $d_{\scriptscriptstyle FL}$
for which the image of the single mode fiber by the lens lies
exactly on the blade. This yields the typical characteristic length $c' = 40~\mu$m reported above in 
Fig.~\ref{couteau}(b). When the refraction effects induced by the Couette cell are taken into account,
the length $c$ displayed in Fig.~\ref{geometryflow} is given by 
$c = f_x c'/\sin(f_\theta \theta_i) \approx 100~\mu$m. 

Using the same method with laser $La_{\scriptscriptstyle 1}$,
we obtained the transverse dimension $a$ of the scattering volume: $a \approx 20~\mu$m.
This characteristic size is consistent with the relation
for Gaussian beams: $ a \approx \lambda f / (\pi\omega_{\scriptscriptstyle{0}}) \approx 20~\mu$m where 
$\omega_{\scriptscriptstyle{0}} \approx 1$~mm is the 
beam waist of the incident light and $f=10$~cm is the focal length of lens $L$.

\subsubsection{Numerical integration of $C(\tau)$}

\begin{figure}[htbp]
\begin{center}
\scalebox{0.4}{\includegraphics{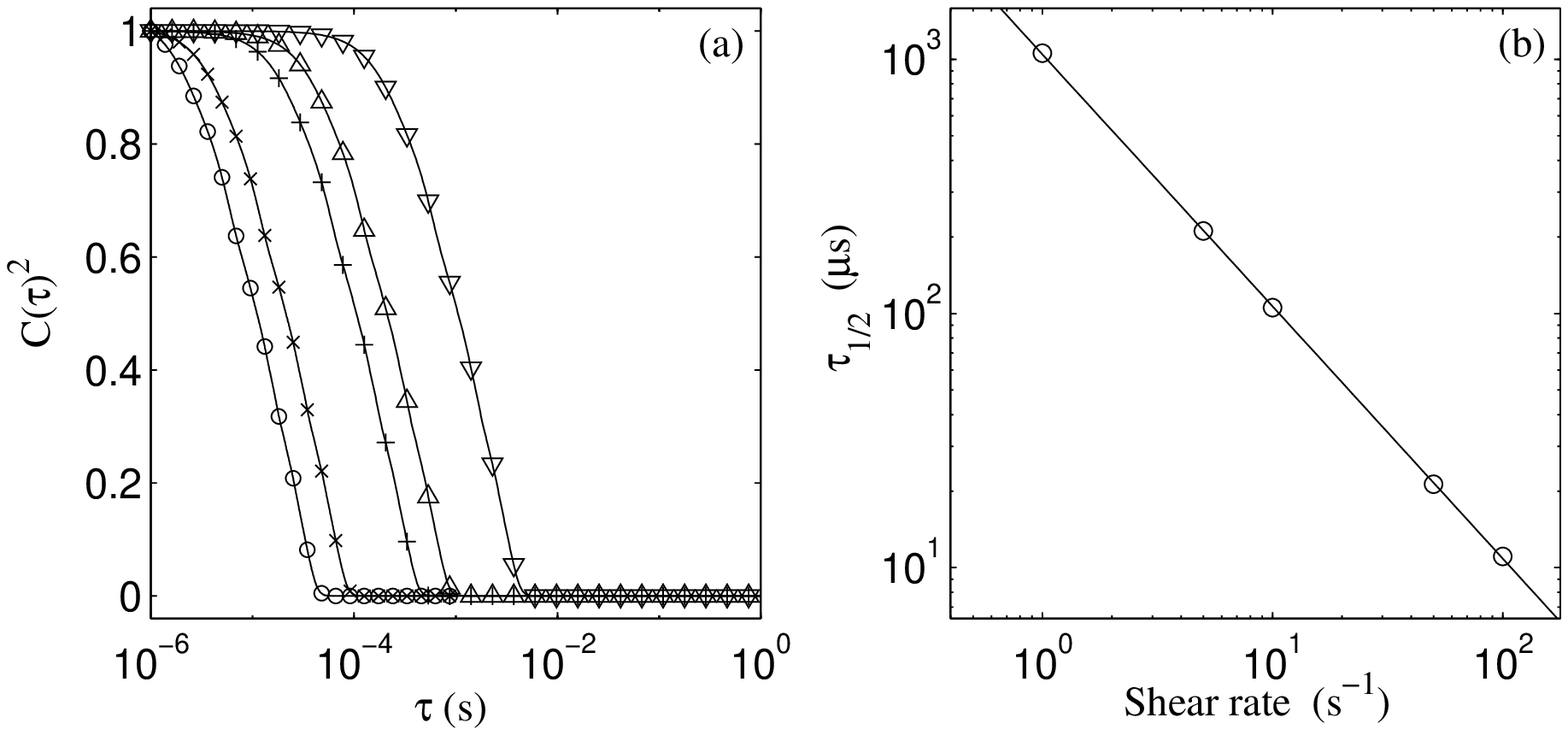}}
\end{center}
\caption{\label{numerique}
(a) Theoretical homodyne correlation functions computed from Eq.~(\ref{tf_volume_diffusant}) with $v=0$ 
for various shear rates  $\gammap = 1$~s$^{-1}~(\triangledown)$, 5~s$^{-1}~(\triangle)$, 10~s$^{-1}~(+)$, 
50~s$^{-1}~(\times)$ and 100~s$^{-1}~(\circ)$.
(b) Computed $\tau_{\scriptscriptstyle{1/2}}$ vs. shear rate. The solid line is the best fit by a power law:
$\tau_{\scriptscriptstyle{1/2}} \propto \gammap^{-0.99}$}
\end{figure}

We computed $C(\tau)$ from Eq.~(\ref{tf_volume_diffusant}) using the experimental 
profile for the scattering volume and with a space dilation factor $f_x/\sin\theta$ due to the Couette cell. 
This leads to the results displayed in Fig.~\ref{numerique}.    
In Fig.~\ref{numerique}(a) is shown the square of heterodyne correlation functions 
calculated from Eq.~(\ref{tf_volume_diffusant}) with $v=0$ for 
various shear rates. These functions are thus equal to
$|g^{(1)}(\tau)|^2$ and are also
exactly equal to the normalized homodyne correlation functions (see Eq.~(\ref{g1tau_general})). 
In Fig.~\ref{numerique}(b), we plotted the half-time $\tau_{\scriptscriptstyle{1/2}}$ 
of the computed homodyne functions,
defined by $C(\tau_{\scriptscriptstyle{1/2}}) = 1/2$, 
vs. various shear rates. All the qualitative features of  homodyne DLS under shear
are reproduced by Eq.~(\ref{tf_volume_diffusant}).
For instance, the continuous line in Fig.~\ref{numerique}(b) shows that
$\tau_{\scriptscriptstyle{1/2}} \propto \gammap^{-1}$, as expected from the physical ideas 
discussed above.

\subsection{Calibration of the experiment using a Newtonian fluid}

To proceed further, one has to perform calibration experiments with a well-known flow.
We used a dilute suspension of latex spheres ($\phi \approx 0.1~\%$~wt.) in a 1:1 water-glycerol
mixture. The optical index of the suspension was measured to be $n=1.40\pm 0.01$.
The imposed scattering angle is $\theta_i = 35^\circ$.
This mixture is a Newtonian solution and, in a Couette cell of aspect ratio 
$e/R_{\scriptscriptstyle 2} = 0.12$,
we expect to measure velocity profiles
that are very close to straight lines.
In the following, we present experimental data showing that, even if in principle the homodyne geometry allows
us to estimate the local shear rate, it is not a suitable method to measure velocity profiles
in a Couette cell because of the strong influence of the shape of the scattering volume and because 
of the uncertainties on the exact positions of the rotor and the stator.
On the other hand, we show that the heterodyne geometry is a time- and space-resolved
method to perform local velocimetry. 

\begin{figure}[htbp]
\begin{center}
\scalebox{0.7}{\includegraphics{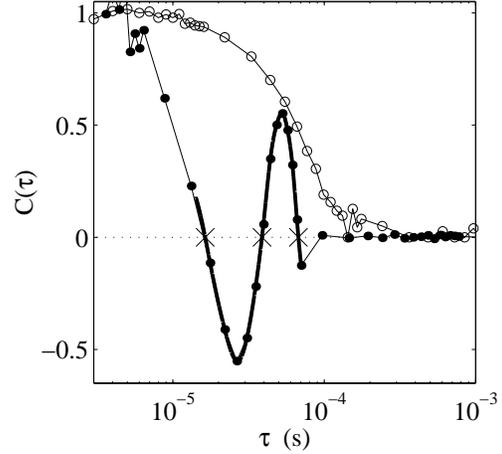}}
\end{center}
\caption{Experimental homodyne ($\circ$)
and heterodyne ($\bullet$) correlation functions recorded on a
latex suspension under shear at $\gammap=10$~s$^{-1}$. The measurements were taken at
$x_t=14.60$~mm {\it i.e.} 1.20~mm away from the moving inner wall.
The thick line shows the portion of the heterodyne
function that is used in the data analysis to determine
the zero crossings ($\times$) from which the local velocity
is calculated (see text).}
\label{exphomohete}
\end{figure}

Typical normalized correlation functions obtained under shear ($\gammap =10$~s$^{-1}$) 
are presented in Fig.~\ref{exphomohete} where the square root of the homodyne
function is plotted in order to be compared to the heterodyne one.
As expected from Eqs.~(\ref{chomo}) and (\ref{g1tau_general}), $C(\tau)^{1/2}$ in homodyne mode
corresponds to the envelope of the oscillating heterodyne function.
Moreover this envelope shows a strong 
decorrelation at time $\tau \approx 80~\mu$s.

\subsubsection{Experimental features of homodyne DLS under shear}
\label{s.expfeat}

\begin{figure}
\begin{center}
\scalebox{0.7}{\includegraphics{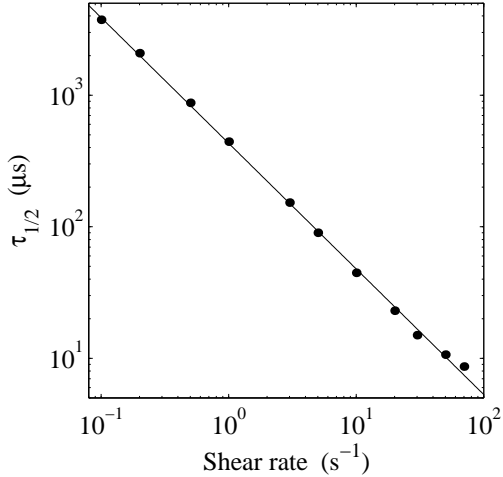}}
\end{center}
\caption{Half-time $\tau_{\scriptscriptstyle{1/2}}$ of the homodyne correlation
function measured in the middle of the gap
for $d_{FL}=200~\mu$m vs. shear rate.
For each data point, the correlation function was
accumulated over 1~min.
The solid line is a power-law fit with exponent
-0.99.}
\label{tps_cis}
\end{figure}

Let us first consider \textit{homodyne} correlation functions. In Fig.~\ref{tps_cis}, the half-time 
$\tau_{\scriptscriptstyle{1/2}}$ of homodyne correlation functions is plotted against the imposed shear rate 
$\gammap$. This time scales as $\tau_{\scriptscriptstyle{1/2}} \propto \gammap^{-1} $.  
Since the characteristic geometrical times 
scale as $\tau_{\gammap} \propto \gammap^{-1}$
and $\tau_{\mathcal{A}} \propto \gammap^{-1}$,
this is the scaling expected from theory.
Moreover the measured $\tau_{\scriptscriptstyle{1/2}}(\gammap)$ are in good agreement with those
computed from Eq.~(\ref{tf_volume_diffusant}) and displayed in Fig.~\ref{numerique}(b).

\begin{figure}[htbp]
\begin{center}
\scalebox{0.7}{\includegraphics{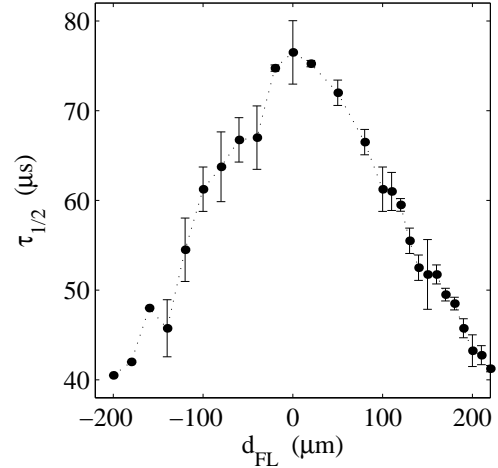}}
\end{center}
\caption{Half-time $\tau_{\scriptscriptstyle{1/2}}$ of the homodyne correlation
function measured in the middle of the gap
at $\gammap=10$~s$^{-1}$ vs. ``defocusing'' distance $d_{FL}$.
For each data point, two correlation functions were
accumulated over 1~min and the two values of $\tau_{\scriptscriptstyle{1/2}}$ were averaged.
The error bars account for the
difference between those two estimates.}
\label{homodyne_defocusing}
\end{figure}

In Fig.~\ref{homodyne_defocusing}, 
we plotted $\tau_{\scriptscriptstyle{1/2}}$ for different defocusing distances $d_{\scriptscriptstyle FL}$ at a given
shear rate $\gammap = 10$~s$^{-1}$.
As explained above, changing the distance $d_{\scriptscriptstyle FL}$ between the lens and the single mode fiber
increases 
the width of the beam at a fixed distance $R$. 
Therefore one expects an increase of the size of the scattering volume 
$\mathcal{V}$. As shown in Fig.~\ref{homodyne_defocusing}, this leads to a decrease of the 
characteristic time $\tau_{\scriptscriptstyle{1/2}} $. This behaviour can be
expected from theory if one considers $\tau_{\gammap} \propto (c\gammap)^{-1}$.
From these results, it seems that the relevant time which governs the correlation
function $C(\tau)$ is $\tau_{\gammap}$. It is in agreement with the features 
discussed in Sec.~\ref{s.numerical_integration}, where we emphasized that 
$\tau_{\mathcal{A}}$ should not play any significant role thanks to the use of single mode fibers.

Homodyne DLS allows quantitative measurements of the local shear rate
only if the shape of the scattering volume $\mathcal{V}$ is precisely known.
In our experiment, the 
scattering volume is obviously not well defined mainly because of the refraction effects
induced by the Couette cell.
For instance a small motion the Couette cell can
change the shape of the scattering volume.

\begin{figure}[htbp]
\begin{center}
\scalebox{0.7}{\includegraphics{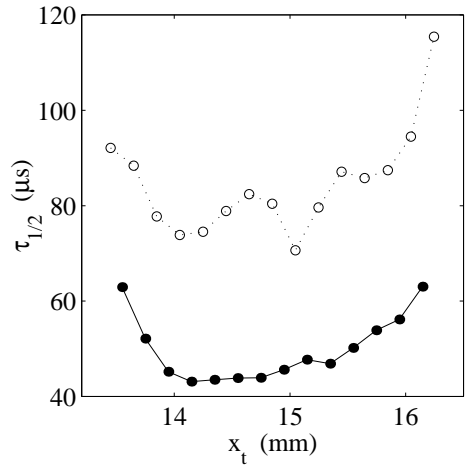}}
\end{center}
\caption{Half-time $\tau_{\scriptscriptstyle{1/2}}$ of the homodyne correlation
function vs. position in the gap for a well-focussed collection
optics $d_{FL}=0$ ($\circ$) and for a
defocussed case $d_{FL}\simeq 200~\mu$m
($\bullet$).
For each data point, the correlation function was
accumulated over 1~min. The shear rate was set to
$\gammap=10$~s$^{-1}$.}
\label{homodynegap}
\end{figure}

In Fig.~\ref{homodynegap},
we plotted the half-time $\tau_{\scriptscriptstyle{1/2}}$ of homodyne correlation functions for various
positions $x_t$ of the mechanical table and for two different defocusing distances 
$d_{\scriptscriptstyle FL}$.
For $d_{\scriptscriptstyle FL} = 0$, the values of $\tau_{\scriptscriptstyle{1/2}}$ show 
a dispersion of about $20~\%$ and for $d_{\scriptscriptstyle FL} = 200~\mu$m,
the dispersion is about $50~\%$. In both cases
the half-times significantly grow when approaching the two walls. 
This effect, which accounts for most of the observed dispersion, is due to the fact that the size 
of scattering volume decreases when $\mathcal{V}$ intersects one of the cell walls.
This increase of $\tau_{\scriptscriptstyle{1/2}}$ is particularly noticeable when 
$d_{\scriptscriptstyle FL} = 200~\mu$m because in this case the scattering volume 
is about twice as large as for $d_{\scriptscriptstyle FL} = 0$ (see Fig.~\ref{couteau2}).  

These intricacies induced by the Couette geometry do not allow for simple quantitative
measurements of the local shear rate.
Moreover  with such measurements, we are not able to determine the exact positions of the 
rotor and the stator of the Couette cell. 
These are the reasons why we chose heterodyne DLS for local measurements under shear.

\subsubsection{Experimental features of heterodyne DLS under shear}
\label{s.velocity.profiles.latex}

In order to extract the Doppler shift $\mathbf{q}\cdot\mathbf{v}$ from the experimental
\textit{heterodyne} correlation function, 
we interpolate the experimental
curves as shown in Fig.~\ref{exphomohete}.
A series of cancellation times  $ \tau_k$ is extracted from the interpolation. The $ \tau_k$'s are
marked by $\times$ symbols in Fig.~\ref{exphomohete}. 
Since $C(\tau)\propto F(\tau)\,\cos(\mathbf{q}\cdot\mathbf{v} \tau)$, one expects $\tau_k = (k\pi + \frac{\pi}{2})(\mathbf{q}\cdot\mathbf{v})^{-1}$ provided $F(\tau)$ does not vanish.
It is then straightforward to estimate the scalar product $\mathbf{q}\!\cdot\!\mathbf{v}$
from a linear fit
of $ \tau_k$ vs. $k$.

\begin{figure}[htbp]
\begin{center}
\scalebox{0.4}{\includegraphics{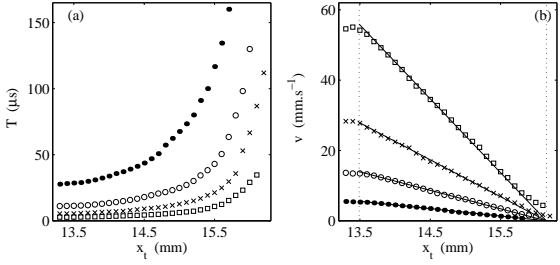}}
\end{center}
\caption{Calibration of the velocity
measurements using a Newtonian suspension of latex spheres
in a water-glycerol mixture 
submitted to different shear rates
$\gammap=2$~s$^{-1}~(\bullet)$, 5~s$^{-1}~(\circ)$, 10~s$^{-1}~(\times)$,
and 20~s$^{-1}~(\Box)$.
(a) Raw estimates of the period $T$ of the oscillations in the
heterodyne correlation function as a function of the position
of the mechanical table $x_t$. The origin of $x_t$ is arbitrary
and is defined when the actuators are first initialized.
(b) Velocity estimates using the procedure described in Sec.~\ref{s.velocity.profiles.latex}. Each data set is fitted
by a straight line, which yields an estimate for the position
of the fixed outer wall: $x_f=16.15\pm 0.02$~mm (dotted line). The dotted line at
$x_m=13.49$~mm shows the estimated
position of the moving inner wall.
The solid lines are the linear velocity profiles
$v=(R_{\scriptscriptstyle 1}\gammap/F_{\gammap}) \,(x_f-x_t)/(x_f-x_m)$
with $\gammap=2$, 5, 10,
and 20~s$^{-1}$.}
\label{calibration}
\end{figure}

To obtain velocity profiles, one has to move the scattering volume across the gap of the Couette cell.
This is done by moving the rheometer in the $x$ direction
with the help of the mechanical table (cf. Fig.~\ref{setup}).
The Doppler period $T = (\mathbf{q}\!\cdot\!\mathbf{v})^{-1}$ is plotted for 
different positions $x_t$ of the table and for different imposed shear rates in Fig.~\ref{calibration}(a).
To estimate the velocity corresponding to a given Doppler period, one has to know precisely the 
scattering vector $\mathbf{q}$. As explained in Sec.~\ref{refraction} refraction 
leads to important optical corrections.
Indeed the real scattering angle $\theta$ differs from the imposed angle    
$\theta_i$ since $\theta \approx f_\theta \theta_i$, where
$f_\theta \approx 0.79$.
Such a correction allows us to convert the measured period $T$ into a velocity $v$
as shown in Fig.~\ref{calibration}(b). Indeed, assuming that $\mathbf{v} = v \, \mathbf{e}_y$,
i.e. that the flow is purely orthoradial in the $(x,y)$ plane, 
$\mathbf{q}\!\cdot\!\mathbf{v}=qv\cos(\theta/2) = 2\pi n\sin\theta\,v/\lambda$.

Linear fits of the measured velocity profiles allow us to estimate the position of the 
fixed outer wall $x_f=16.15$~mm with a precision of about $20~\mu$m. The velocity $v_{\scriptscriptstyle 0}$
of the rotor is given by the
rheometer: for a given angular velocity $\Omega$ of the rotor, 
the shear rate $\gammap$ indicated by the rheometer, $\gammap = F_{\gammap}\Omega$ where 
$F_{\gammap} = (R_{\scriptscriptstyle 1}^2 + R_{\scriptscriptstyle 2}^2)/(R_{\scriptscriptstyle 2}^2 - R_{\scriptscriptstyle 1}^2)$ with $R_{\scriptscriptstyle 1}$ and $R_{\scriptscriptstyle 2}$ denoting the radii of the 
rotor and the stator of the Couette cell. $F_{\gammap}$ is a conversion factor inherent to the rheometer 
that account for the Couette geometry.
The velocity of the mobile inner wall is thus given by: 
$v_{\scriptscriptstyle 0} = R_{\scriptscriptstyle 1}\Omega = 
R_{\scriptscriptstyle 1}\gammap/F_{\gammap}$.  
This permits the determination of the position of the moving inner wall $x_m=13.49$~mm
(left dashed line in Fig.~\ref{calibration}(b)). 
Such a determination leads to an effective gap width $e'=x_f-x_m\approx 2.65$~mm, 
which differs from the real gap $e = 3$~mm. 
Actually as shown in Sec.~\ref{refraction}, the actual displacement
of the scattering volume inside the cell $\delta x'$ is related to the displacement
of the mechanical table $\delta x$,
by the following relation: $\delta x' = f_x \delta x$, where $f_x \approx 1.13$
is deduced from the Snell-Descartes law for refraction. Such a conversion factor is in
excellent agreement with the previous result since $e/e' \approx 1.13$.
Let us notice that for all the imposed shear rates, we were able to measure velocities 
for a few points that are outside the gap at $x_t < x_m$ (see Fig.~\ref{calibration}(b)). 
This is due to the finite extent of the scattering volume: even for $x_t < x_m$, some
small intensity may be detected that is scattered from the intersection of the scattering volume and 
the gap.

\begin{figure}[htbp]
\begin{center}
\scalebox{0.7}{\includegraphics{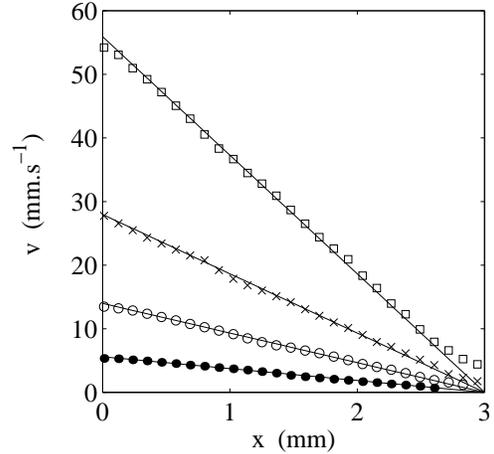}}
\end{center}
\caption{
Velocity profiles in a sheared Newtonian suspension
for $\gammap=2$~s$^{-1}~(\bullet)$, 5~s$^{-1}~(\circ)$, 10~s$^{-1}~(\times)$,
and 20~s$^{-1}~(\Box)$.
The abscissa $x$ denotes the real position within
the cell gap. The origin of $x$ is taken at the inner wall.
The solid lines are the linear velocity profiles
$v=R_{\scriptscriptstyle 1}\gammap(1-x/e)/F_{\gammap}$
with $\gammap=2$, 5, 10,
and 20~s$^{-1}$. 
The conversion factors $f_\theta=0.79$ and $f_x=1.13$ are used to account
for refraction effects}
\label{profillatex}
\end{figure}

In the velocity profiles displayed in Fig.~\ref{profillatex},
the two optical conversion factors $f_\theta$ and $f_x$ have been taken into account and $x$ denotes 
the real position inside the gap.
In all cases, the uncertainty of the experimental measurements is of the order of the marker size. 
Note that the effect of non-homogeneous shear rate due to
the curved geometry of the Couette cell
is negligible for such a Newtonian fluid and does not influence the calibration.

\subsubsection{Spatial resolution}

One of the most relevant features of a technique measuring velocity profiles is its spatial resolution.
As previously outlined, a good resolution  would be around $100~\mu$m.
In our case, spatial resolution is limited by the size of the scattering volume $\mathcal{V}$. Indeed,
velocity measurements using heterodyne DLS are averaged over $\mathcal{V}$.
The typical dimensions of $\mathcal{V}$ in air are $a \approx 20~\mu$m and $c' \approx 40~\mu$m measured in
Sec.~\ref{s.size}. 
Since scattering occurs at an angle $\theta$ and since the Couette cell acts as a lens, 
the relevant size is $c \approx f_x c'/ \sin( f_{\theta} \theta_i)$, which leads
to a typical size $c \approx 100~\mu$m.  
However, the spatial resolution of our setup strongly depends on the scattering angle. The latest  
measurements performed in our group on banded flows in micellar solution at 
a larger angle ($\theta_i \approx 50^\circ$) 
reveal that our setup can resolve
a discontinuity of the shear rate with an accuracy of about $50~\mu$m \cite{Molino:02}. 

\subsubsection{Temporal resolution}

As recalled in the Introduction, some complex fluids display non-stationnary flows 
\cite{Bandyopadhyay:01,Salmon:02}.
Therefore it is interesting to perform time-resolved measurements of the local velocity. 
To obtain the Doppler frequency with a high temporal resolution
one may use a spectrum analyzer rather than a correlator. 
Such a technique allows up
to a thousand measurements per second. For instance, this has been 
implemented long ago by Gollub \textit{et al.} to 
measure the temporal evolution of the local velocity in Newtonian fluids 
at the onset of the Taylor-Couette instability
\cite{Gollub:75}.
However the use of a correlator yields more information than a spectrum analyzer. In homodyne mode
qualitative information on the local shear rate can be obtained, 
as discussed in Secs.~\ref{s.theory} and \ref{s.expfeat}. Moreover, the use of a correlator 
once the shear flow has been stopped can give important information on the coupling between 
the structural dynamics of a material and an applied shear \cite{Viasnoff:02}.

\begin{figure}[htbp]
\begin{center}
\scalebox{0.4}{\includegraphics{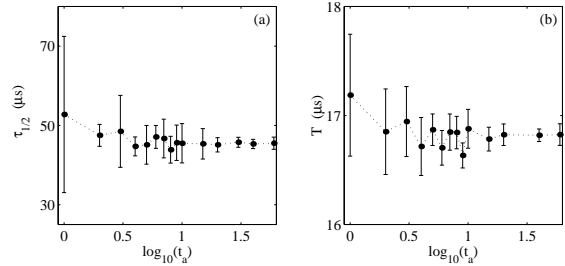}}
\caption{Statistical convergence
of the estimates of the characteristic times.
(a) Half-time $\tau_{1/2}$ of the homodyne correlation
function vs. accumulation time $t_a$. (b) Period $T$ of the oscillations
in the heterodyne correlation function vs. accumulation time $t_a$.
Each data point represents the average over the
the characteristic times
estimated from five consecutive correlation functions
accumulated over $t_a$. The error bars account for the
standard deviation of those five measurements.
The shear rate was set to $\gammap=10$~s$^{-1}$.}
\label{asymptotisation}
\end{center}
\end{figure}

The temporal resolution of our setup strongly depends on the power of the laser beam and
on the quantity of light scattered by the sample. 
Actually these two features control $N(t)$, the number of photons
detected by the photomultiplier tube per second. To estimate the temporal resolution of our setup, 
we measured the homodyne and heterodyne correlation functions 
for various accumulation times $t_a$ at a given
$N(t)$. Then we computed the corresponding half-times of the  homodyne correlation functions 
and the Doppler periods in the heterodyne mode.
Figure~\ref{asymptotisation} displays such measurements obtained on a latex suspension in the middle of 
the gap for $\gammap = 10$~s$^{-1}$ and with $N(t) \approx 10^4$~events per second in homodyne mode.
The error bars represent the standard deviations of different measurements 
obtained at a given accumulation time $t_a$. It clearly appears that good statistical estimates of 
$\tau_{\scriptscriptstyle{1/2}}$ and $T = (\mathbf{q}\!\cdot\!\mathbf{v})^{-1}$
are obtained after an accumulation time of the order of $5$~seconds.
Actually in better experimental conditions, for instance when $N(t) \approx 3.10^4$, 
the accumulation time required to
define those characteristic times decreases to $1$~second. 
Therefore our setup is well suited to follow  
dynamics like those observed in Refs.~\cite{Bandyopadhyay:01,Salmon:02} that 
involve time scales in the range 1--$10^3$~s. 

\section{Discussion and conclusion}

In this article, we presented an experimental setup based on the use of single mode fibers that allows
one to measure velocity profiles in Couette flows.
We have shown that measurements of the local shear rate 
using homodyne DLS are difficult to implement in the Couette geometry.
By measuring directly the shape of the scattering volume, 
we were able to compare experimental and theoretical homodyne correlation
functions, and show that the decorrelation times $\tau_{\scriptscriptstyle{1/2}}$ 
are linked to the size of the scattering volume $c$ and to the local shear rate  $\gammap$ according to 
$\tau_{\scriptscriptstyle{1/2}} \propto 1/(c \gammap)$ as expected from theory. 

Heterodyne detection yields the Doppler shift $\mathbf{q}\cdot\mathbf{v}$ associated to the average velocity in the 
scattering volume. Provided refraction effects are taken into account and a careful calibration is performed,
our setup  gives access to velocity profiles in 
Couette flows of complex fluids down to a spatial resolution of 50~$\mu$m.
As far as temporal resolution is concerned, the major limitation is the time 
required to move the rheometer along the direction of shear. In the best experimental conditions
and with about $30$ points in the gap of the Couette cell, 
we managed to measure velocity profiles in about one minute.
Thus, our setup cannot resolve flow dynamics over the whole gap on time scales 
shorter than one minute. 
Other non intrusive techniques like ultrasonic speckle correlation recently developed in our group may allow us  
to measure up to $10^2$ velocity profiles per second and may yield rich information
on faster dynamics.      

\begin{figure}[htbp]
\begin{center}
\scalebox{0.4}{\includegraphics{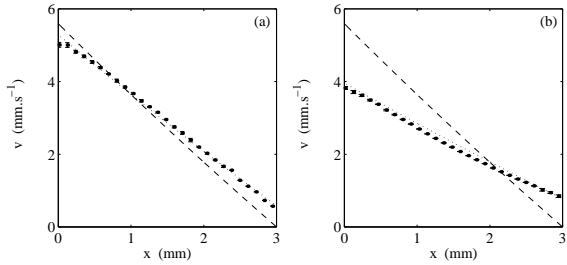}}
\caption{Velocity profiles for  $\gammap=2$~s$^{-1}$ in two different emulsions. 
(a) Dilute emulsion $\phi = 20~\%$. (b) Concentrated emulsion $\phi = 75~\%$. 
The dashed lines represent the velocity profiles expected for a Newtonian fluid.
The dotted lines correspond to a Newtonian fluid when wall
slip has been taken into account}
\label{optique_emulsion}
\end{center}
\end{figure}

In order to show its relevance in the field of complex fluid flows,
the DLS technique described in this paper was applied to \textit{sheared emulsions}.
Figure~\ref{optique_emulsion} shows the velocity
profiles measured in two different oil-in-water emulsions at the same shear rate.
The Couette cell is that used throughout this study ($e=3$~mm and $R_{\scriptscriptstyle 1}=22$~mm).
Both emulsions are composed of silicone oil droplets
(diameter $\approx 2~\mu$m) in a water--glycerol mixture with 1~\% TTAB surfactant.
Figure~\ref{optique_emulsion}(a) corresponds to a dilute emulsion of volume fraction
$\phi = 20~\%$ whereas Fig.~\ref{optique_emulsion}(b) was obtained with a concentrated
system of volume fraction $\phi = 75~\%$. For $\gammap=2$~s$^{-1}$,
rheological data recorded by the rheometer show that the global
stresses imposed on the samples are $\sigma=0.07$~Pa and $\sigma=290$~Pa respectively.
The optical index $n$ of the aqueous phase is matched to that of the silicone oil to avoid multiple
scattering. Moreover, $n=1.40$ as in the latex suspension used for the calibration,
so that the values of $f_x$ and $f_\theta$ discussed in Sec.~\ref{refraction}
and \ref{s.velocity.profiles.latex} remain valid in the case of those emulsions.
In both cases, slippage clearly shows on the velocity profiles when compared to that of a Newtonian
fluid. Wall slip is much larger for the concentrated system. Moreover, the velocity profile
is almost perfectly linear in the dilute emulsion (see fig.~\ref{optique_emulsion}(a))
whereas a small but systematic curvature can be noticed in the concentrated
emulsion (see fig.~\ref{optique_emulsion}(b)). This curvature is the signature of the
non-Newtonian behavior of the 75~\% emulsion. The complete analysis of the slip velocities and
of the bulk flow behavior inferred from such local measurements will
be discussed at length in a related publication \cite{Salmon:02_3}.

Finally, the main innovation of the present work is to provide both local velocity measurements and
global rheological data simultaneously.
DLS coupled to standard rheology opens the door to \textit{local
rheology}. Indeed, when the flow is stationnary,
the local stress in a Couette cell is inversely proportional to the square of the position in the gap. 
On the other hand,
the local shear rate may be accessed by differentiating the velocity profiles.
Thus, we could determine the relationship between the local shear rate and the
local stress and define a local flow behaviour. The comparison between such a local rheology
and the global flow curve obtained with a standard rheometer should lead to essential knowledge
that may prove crucial to understand complex fluid flows \cite{Fisher:01,Salmon:02_3,Molino:02,Coussot:02}.

\begin{acknowledgement}
The authors are deeply grateful to the ``Cellule Instrumentation'' at CRPP for
building the heterodyne DLS setup and the Couette cells used in this study.
We also thank F. Nadal for technical help on the figures of this paper.
\end{acknowledgement}

\end{document}